\documentclass[superscriptaddress, aps, prl, 10pt, amsmath, amssymb, bibnotes,
altaffilletter, twocolumn, floatfix]{revtex4-2}

\usepackage{hyperref}
\usepackage[T1]{fontenc}
\usepackage{lineno}
\usepackage{xcolor}
\usepackage{graphicx} 
\graphicspath{{/}}
\usepackage{dcolumn} 
\usepackage{bm} 
\hypersetup{colorlinks=true, citecolor=blue, urlcolor=blue, linkcolor=blue}
\usepackage{orcidlink}

\def\MJ{{\sc Majorana}}
\def\DEM{{\sc Demonstrator}}
\def\MJD{{\sc Majorana Demonstrator}}
\def\nonubb{$\beta\beta(0\nu)$}
\def\enrge{${}^{\mathrm{enr}}$Ge}
\def\natge{${}^{\mathrm{nat}}$Ge}
\def\gae{$g_{ae}$}

\begin{document}


\title{Exotic dark matter search with the \textsc{Majorana Demonstrator}}



\newcommand{\uw}{Center for Experimental Nuclear Physics and Astrophysics, and Department of Physics, University of Washington, Seattle, WA 98195, USA}
\newcommand{\ITEP}{National Research Center ``Kurchatov Institute'' Institute for Theoretical and Experimental Physics, Moscow, 117218 Russia}
\newcommand{\JINR}{Joint Institute for Nuclear Research, Dubna, 141980 Russia}
\newcommand{\lbnl}{Nuclear Science Division, Lawrence Berkeley National Laboratory, Berkeley, CA 94720, USA}
\newcommand{\lbnle}{Engineering Division, Lawrence Berkeley National Laboratory, Berkeley, CA 94720, USA}
\newcommand{\lanl}{Los Alamos National Laboratory, Los Alamos, NM 87545, USA}
\newcommand{\queens}{Department of Physics, Engineering Physics and Astronomy, Queen's University, Kingston, ON K7L 3N6, Canada}
\newcommand{\unc}{Department of Physics and Astronomy, University of North Carolina, Chapel Hill, NC 27514, USA}
\newcommand{\duke}{Department of Physics, Duke University, Durham, NC 27708, USA}
\newcommand{\ncsu}{Department of Physics, North Carolina State University, Raleigh, NC 27695, USA}
\newcommand{\ornl}{Oak Ridge National Laboratory, Oak Ridge, TN 37830, USA}
\newcommand{\ou}{Research Center for Nuclear Physics, Osaka University, Ibaraki, Osaka 567-0047, Japan}
\newcommand{\pnnl}{Pacific Northwest National Laboratory, Richland, WA 99354, USA}
\newcommand{\ttu}{Tennessee Tech University, Cookeville, TN 38505, USA}
\newcommand{\sdsmt}{South Dakota Mines, Rapid City, SD 57701, USA}
\newcommand{\usc}{Department of Physics and Astronomy, University of South Carolina, Columbia, SC 29208, USA}
\newcommand{\usd}{Department of Physics, University of South Dakota, Vermillion, SD 57069, USA}
\newcommand{\ut}{Department of Physics and Astronomy, University of Tennessee, Knoxville, TN 37916, USA}
\newcommand{\tunl}{Triangle Universities Nuclear Laboratory, Durham, NC 27708, USA}
\newcommand{\mpi}{Max-Planck-Institut f\"{u}r Physik, M\"{u}nchen, 80805, Germany}
\newcommand{\tum}{Physik Department and Excellence Cluster Universe, Technische Universit\"{a}t, M\"{u}nchen, 85748 Germany}
\newcommand{\williams}{Physics Department, Williams College, Williamstown, MA 01267, USA}
\newcommand{\ciemat}{Centro de Investigaciones Energ\'{e}ticas, Medioambientales y Tecnol\'{o}gicas, CIEMAT 28040, Madrid, Spain}
\newcommand{\iu}{Department of Physics, Indiana University, Bloomington, IN 47405, USA}
\newcommand{\iuceem}{IU Center for Exploration of Energy and Matter, and Department of Physics, Indiana University, Bloomington, IN 47405, USA}

\author{I.J.~Arnquist\,\orcidlink{0000-0002-5643-8330}}\affiliation{\pnnl}
\author{F.T.~Avignone~III}\affiliation{\usc}\affiliation{\ornl}
\author{A.S.~Barabash\,\orcidlink{0000-0002-5130-0922}}\affiliation{\ITEP}
\author{C.J.~Barton}\altaffiliation{Present address: Roma Tre University and INFN Roma Tre, Rome, Italy}\affiliation{\usd}
\author{K.H.~Bhimani}\affiliation{\unc}\affiliation{\tunl}
\author{E.~Blalock\,\orcidlink{0000-0001-5311-371X}}\affiliation{\ncsu}\affiliation{\tunl}
\author{B.~Bos}\affiliation{\unc}\affiliation{\tunl}
\author{M.~Busch}\affiliation{\duke}\affiliation{\tunl}
\author{M.~Buuck\,\orcidlink{0000-0001-5751-4326}}\affiliation{\uw}
\author{T.S.~Caldwell}\affiliation{\unc}
\author{Y-D.~Chan}\affiliation{\lbnl}
\author{C.D.~Christofferson}\affiliation{\sdsmt}
\author{P.-H.~Chu\,\orcidlink{0000-0003-1372-2910}}\affiliation{\lanl}
\author{M.L.~Clark}\affiliation{\unc}\affiliation{\tunl}
\author{C.~Cuesta\,\orcidlink{0000-0003-1190-7233}}\affiliation{\ciemat}
\author{J.A.~Detwiler\,\orcidlink{0000-0002-9050-4610}}\affiliation{\uw}
\author{Yu.~Efremenko}\affiliation{\ut}\affiliation{\ornl}
\author{H.~Ejiri}\affiliation{\ou}
\author{S.R.~Elliott\,\orcidlink{0000-0001-9361-9870}}\affiliation{\lanl}
\author{G.K.~Giovanetti}\affiliation{\williams}
\author{M.P.~Green\,\orcidlink{0000-0002-1958-8030}}\affiliation{\ncsu}\affiliation{\tunl}\affiliation{\ornl}
\author{J.~Gruszko\,\orcidlink{0000-0002-3777-2237}}\affiliation{\unc}\affiliation{\tunl}
\author{I.S.~Guinn\,\orcidlink{0000-0002-2424-3272}}\affiliation{\ornl}
\author{V.E.~Guiseppe\,\orcidlink{0000-0002-0078-7101}}\affiliation{\ornl}
\author{C.R.~Haufe}\affiliation{\unc}\affiliation{\tunl}
\author{R.~Henning}\affiliation{\unc}\affiliation{\tunl}
\author{D.~Hervas~Aguilar}\affiliation{\unc}\affiliation{\tunl}
\author{E.W.~Hoppe\,\orcidlink{0000-0002-8171-7323}}\affiliation{\pnnl}
\author{A.~Hostiuc}\affiliation{\uw}
\author{M.F.~Kidd}\affiliation{\ttu}
\author{I.~Kim}\altaffiliation{Corresponding author}~\email{inwookkim.physics@gmail.com}\altaffiliation{Present address: Lawrence Livermore National Laboratory, Livermore CA 94550}\affiliation{\lanl}
\author{R.T.~Kouzes}\affiliation{\pnnl}
\author{T.E.~Lannen~V}\affiliation{\usc}
\author{A.~Li\,\orcidlink{0000-0002-4844-9339}}\altaffiliation{Present address: University of California San Diego, La Jolla, CA 92093, USA}\affiliation{\unc}\affiliation{\tunl}
\author{A.M.~Lopez}\affiliation{\ut}
\author{J.M. L\'opez-Casta\~no}\affiliation{\ornl}
\author{E.L.~Martin}\altaffiliation{Present address: Duke University, Durham, NC 27708}\affiliation{\unc}\affiliation{\tunl}	
\author{R.D.~Martin}\affiliation{\queens}
\author{R.~Massarczyk\,\orcidlink{0000-0001-8001-9235}}\affiliation{\lanl}
\author{S.J.~Meijer\,\orcidlink{0000-0002-1366-0361}}\affiliation{\lanl}
\author{S.~Mertens}\affiliation{\mpi}\affiliation{\tum}
\author{T.K.~Oli\,\orcidlink{0000-0001-8857-3716}}\altaffiliation{Present address: Argonne National Laboratory, Lemont, IL 60439, USA}\affiliation{\usd}
\author{G.~Othman}\altaffiliation{Present address: Universit{\"a}t Hamburg, Institut f{\"u}r Experimentalphysik, Hamburg, Germany}\affiliation{\unc}\affiliation{\tunl}
\author{L.S.~Paudel\,\orcidlink{0000-0003-3100-4074}}\affiliation{\usd}
\author{W.~Pettus\,\orcidlink{0000-0003-4947-7400}}\affiliation{\iuceem}
\author{A.W.P.~Poon\,\orcidlink{0000-0003-2684-6402}}\affiliation{\lbnl}
\author{D.C.~Radford}\affiliation{\ornl}
\author{J.~Rager}\altaffiliation{Present address: U.S.~Army DEVCOM Army Research Laboratory Armed Forces, Adelphi, Maryland, 20783, USA}\affiliation{\unc}
\author{A.L.~Reine\,\orcidlink{0000-0002-5900-8299}}\affiliation{\unc}\affiliation{\tunl}
\author{K.~Rielage\,\orcidlink{0000-0002-7392-7152}}\affiliation{\lanl}
\author{N.W.~Ruof\,\orcidlink{0000-0001-9665-6722}}\altaffiliation{Present address: Lawrence Livermore National Laboratory, Livermore, California 94550, USA}\affiliation{\uw}
\author{D.C.~Schaper\,\orcidlink{0000-0002-6219-650X}}\affiliation{\lanl}
\author{D.~Tedeschi}\affiliation{\usc}
\author{R.L.~Varner\,\orcidlink{0000-0002-0477-7488}}\affiliation{\ornl}
\author{S.~Vasilyev}\affiliation{\JINR}
\author{J.F.~Wilkerson\,\orcidlink{0000-0002-0342-0217}}\affiliation{\unc}\affiliation{\tunl}\affiliation{\ornl}
\author{C.~Wiseman\,\orcidlink{0000-0002-4232-1326}}\altaffiliation{Corresponding author}~\email{wisecg@uw.edu}\affiliation{\uw}
\author{W.~Xu}\affiliation{\usd}
\author{C.-H.~Yu\,\orcidlink{0000-0002-0342-0217}}\affiliation{\ornl}
\author{B.X.~Zhu}\altaffiliation{Present address: Jet Propulsion Laboratory, California Institute of Technology, Pasadena, CA 91109, USA}\affiliation{\lanl}

\collaboration{{\sc{Majorana}} Collaboration}
\noaffiliation

\date{December 7, 2023}

\begin{abstract}

  With excellent energy resolution and ultra-low level radiogenic backgrounds, the high-purity germanium detectors in the \MJD\ enable searches for several classes of exotic dark matter (DM) models.
  In this work, we report new experimental limits on keV-scale sterile neutrino DM via the transition magnetic moment from conversion to active neutrinos, $\nu_s \rightarrow \nu_a$.
  We report new limits on fermionic dark matter absorption ($\chi + A \rightarrow \nu + A$) and sub-GeV DM-nucleus 3$\rightarrow$2 scattering ($\chi + \chi + A \rightarrow \phi + A$), and new exclusion limits for bosonic dark matter (axionlike particles and dark photons).
  These searches utilize the (1--100)-keV low energy region of a 37.5-kg y exposure collected by the \DEM\ between May 2016 and November 2019, using a set of $^{76}$Ge-enriched detectors whose surface exposure time was carefully controlled, resulting in extremely low levels of cosmogenic activation.

\end{abstract}


\maketitle

  As large-scale dark matter (DM) experiments have rejected much of the weakly interacting massive particle parameter space, interest in alternative models has increased.
  Popular models include axionlike particles and dark photons~\cite{pospelov2008, gerda2020, xenon_nt_2022}, and fermionic DM~\cite{dror2020absorption, dror2020detecting}. 
  Novel scattering channels for light (sub-GeV) DM have also been proposed~\cite{chao2021_dmnuc32}.
  The \MJD\ conducted a search for neutrinoless double-beta decay [\nonubb] from 2015 to 2019 with a 29.7-kg set of $^{76}$Ge-enriched high-purity germanium (HPGe) $p$-type point contact (PPC) detectors with world-leading energy resolution~\cite{majorana2023final}.
  The unique low-background dataset enables searches for new physics at the keV scale~\cite{wfc2022, solax2022, massarczyk2018, vorren2017}.
  The HPGe detectors routinely achieved $\sim$1-keV energy thresholds, with energy calibration, pulse shape parameters, and analysis thresholds updated with weekly calibration data from two $^{228}$Th sources~\cite{arnquist2023energy}.
  Data were acquired with a statistical blinding scheme, taking cycles of 31 hours open data followed by 93 hours of blind data, interspersed with open calibration runs.
  In 2020 a set of new \enrge\ detectors were installed, bringing the total exposure collected to 65-kg y.
  The \DEM\ continues to operate with 14.3~kg of natural-abundance (\natge) detectors for background studies and new rare-event searches~\cite{ralph2023}.

  \begin{figure}
    \centering
    \includegraphics[width=\columnwidth]{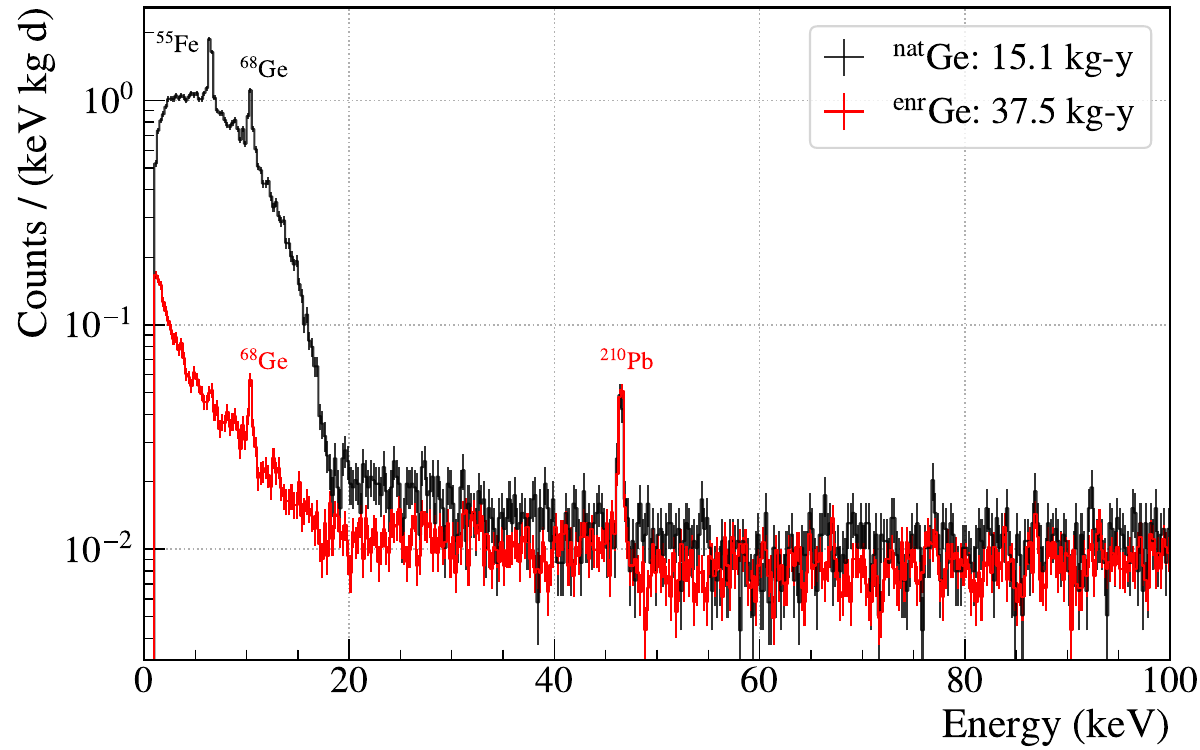}
    \caption{Energy spectra from \MJ\ \enrge\ (red) and \natge\ (black) detectors, 1--100 keV.
    The spectrum in the \natge\ below 20 keV is tritium dominated.
    Lower levels of tritium, $^{55}$Fe, $^{65}$Zn, $^{68}$Ge show that limiting surface exposure of the \enrge\ material significantly reduced cosmogenic activation.
    }
    \label{fig:spec}
  \end{figure}

  From the 2015--2019 data set with the original \enrge\ detectors, a total exposure of 49.05~kg y was collected, and 37.5~kg y was selected for the low-energy analysis, retaining 76\%.
  Detectors with recurring near-threshold electronics noise constitute the majority of the rejected exposure.
  Similarly, 15.1~kg y of \natge\ exposure was selected from 21.97~kg y.
  The \natge\ detectors provided an important cross-check of the analysis and data cleaning routines, but are ultimately not used in this rare event search, due to higher backgrounds at all energies, most notably in the tritium region.
  (This choice was made prior to unblinding based on open data.)
  The energy spectra from both sets of detectors are shown in Fig.~\ref{fig:spec}.

  The voltage-to-energy calibration, time-dependent channel selection, granularity, and muon veto are computed by the \nonubb\ analysis.
  For additional pulse shape discrimination at the lowest energies, waveforms were wavelet denoised and fit to an exponentially modified Gaussian function, whose slope parameter is proportional to the risetime of the full charge collection.
  Energy-degraded $n^+$ surface events have longer risetimes, allowing them to be rejected.
  Details on the slope parameter analysis and tuning are given in Ref.~\cite{wiseman2019}.
  The quoted exposures include a reduction in active mass from this n$^+$ dead layer fiducial volume cut.
  The wavelet coefficients are also used to remove high-frequency noise events with high efficiency.

  \begin{figure}
    \centering
    \includegraphics[width=\columnwidth]{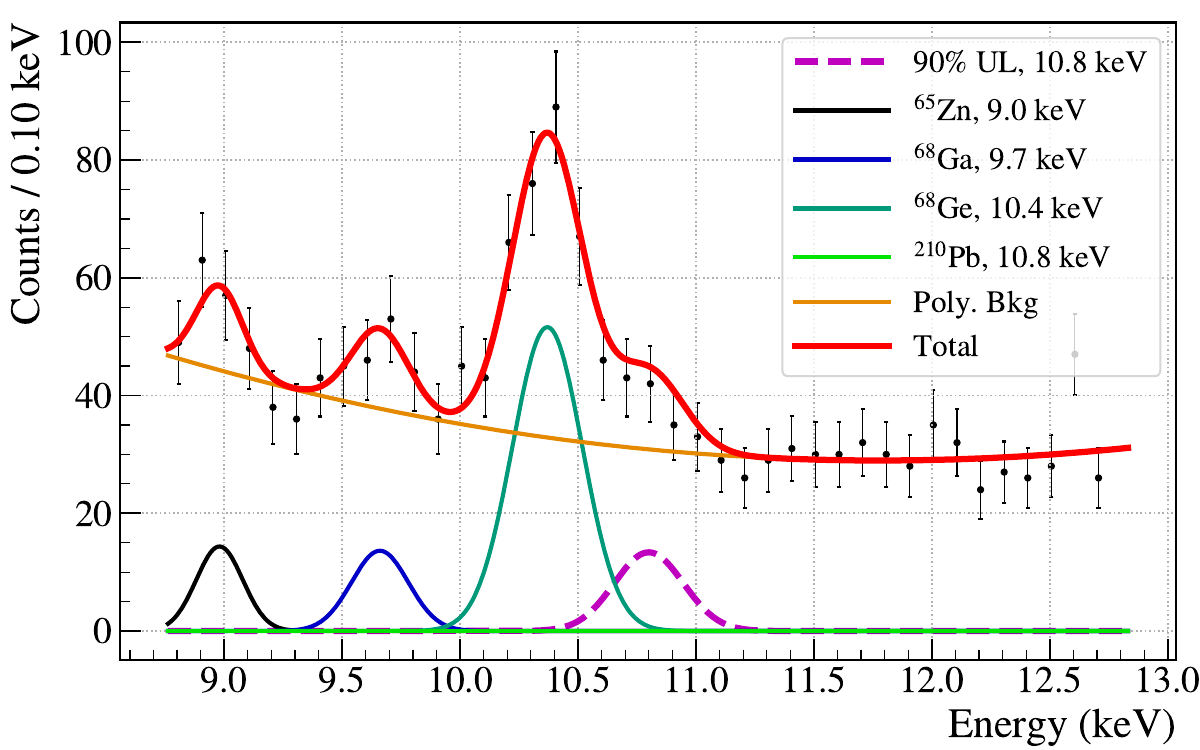}
    \caption{An example fit to the spectrum and calculation of the 90\% upper limit for a rare peak at 10.8~keV. A $^{210}$Pb line is allowed but not observed at this energy.
      The upper limit is obtained from a profile likelihood ratio test.}
    \label{fig:peakfit}
  \end{figure}

\textit{Statistical methods.}--
  We perform a raster scan for a Gaussian peak from a hypothetical rare signal at energy $E_r$ ranging from 1 to 100~keV, sampling intervals at half the expected \MJ\ detector energy resolution $\sigma_{M}$ (keV).
  In this energy range, $\sigma_{M}(E) = (0.138^2 + 0.017^2 E+ 0.00028^2 E^2)^{-1/2}$, weakly increasing from 0.15-keV FWHM at 1~keV to 0.23~keV at 100~keV.
  Each fit is performed in a moving energy window with width $\pm$(7$\sigma_M$ + 1)~keV.
  Our spectral model in each window is given by
  \begin{align}
    \nonumber \frac{dN(E|E_r)}{dE} = & \bigg( n_\mathrm{rare} \mathcal{P}(E|E_r) + \sum_i^{n_\mathrm{pks}} n_i \mathcal{P}(E|E_i) \bigg) \eta(E) \\
    & + b_0 \mathcal{C}_0(E) + b_1 \mathcal{C}_1(E) + b_2  \mathcal{C}_2(E).
  \end{align}
  Here $\mathcal{P}(E|E_k)$ is the (Gaussian) detector response for energy deposition $E_k$, and $n_{\mathrm{rare}}$ and $n_i$ are the observed counts in the hypothetical rare peak and the known background lines, respectively. 
  We include the known long-lived cosmogenic lines from  $^{68}$Ge (1.3 and 10.37 keV), $^{49}$V (4.97), $^{54}$Mn (5.99), $^{55}$Fe (6.54), $^{57}$Co (7.11), $^{65}$Zn (8.98), and $^{68}$Ga (9.66)~\cite{cdms2019}. 
  We also include lines for radiogenic $^{210}$Pb (10.8, 46.5); other radiogenic lines would have contributions at higher energies that are not observed~\cite{buuck2019}.
  The detection efficiency $\eta(E)$ is described in the Appendix, and the sum over Chebyshev polynomials of the first kind $\mathcal{C}_n$ is used to approximate the continuum shape in each fit window.  

\begin{figure}
    \centering
    \includegraphics[width=\columnwidth]{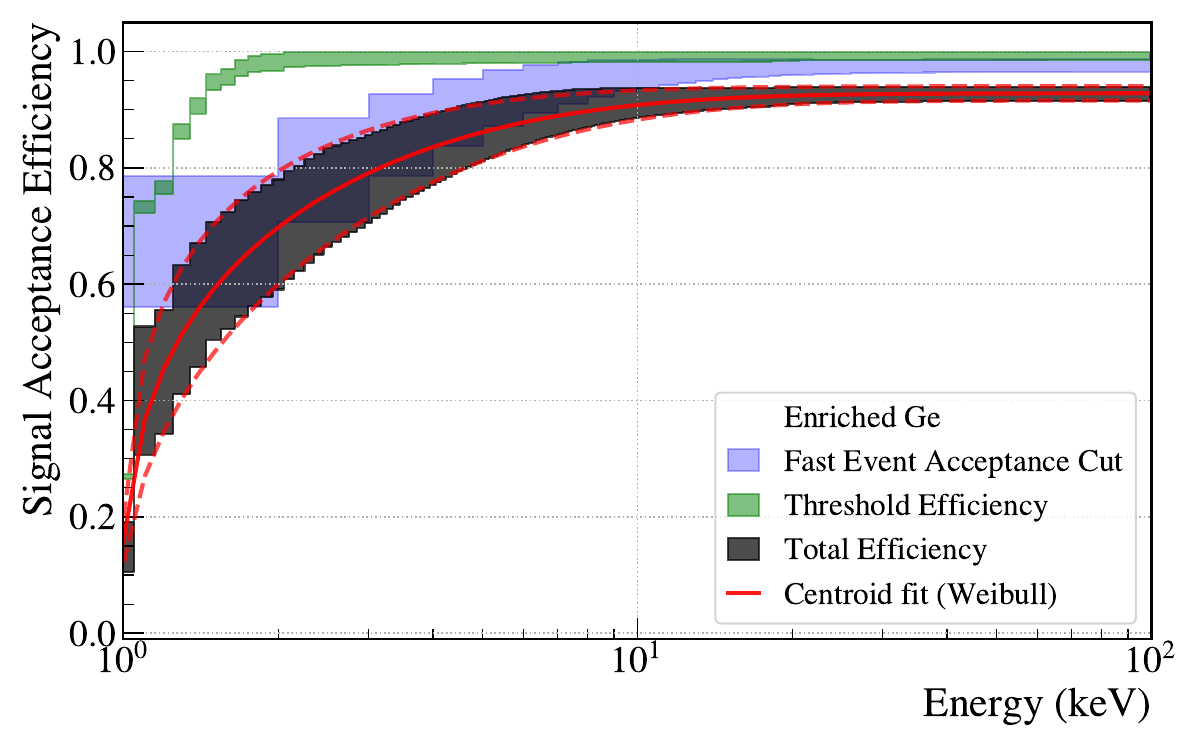}
    \caption{Total efficiency and uncertainty for the \enrge\ detectors used in the rare event search.
    Contributions from varying energy thresholds over time (green) and surface event rejection (blue) are convolved with a flat high-frequency noise rejection efficiency of 95\% to produce the final efficiency (black).
    Shaded regions represent the 1$\sigma$ uncertainty.
    The total and upper and lower bounds are fit with a cumulative Weibull distribution (red).}
    \label{fig:cuteff}
  \end{figure}

  We perform an unbinned extended maximum likelihood fit, with Gaussian constraint terms for the energy resolution, pulse shape cut efficiency, and its uncertainty (Fig.~\ref{fig:cuteff}).
  An example is shown in Fig.~\ref{fig:peakfit}.
  We include a 30\% multiplicative uncertainty in the expected resolution $\sigma_M(E)$ based on studies of background and $^{228}$Th calibration peak widths.
  To test for the presence of a signal, we use the standard profile likelihood ratio as the test statistic~[Eq.~4 in Ref.~\cite{baxter2021recommended}].
  Expected spectral lines with unconstrained yields are included explicitly in this step for $^{55}$Fe (6.54 keV), $^{68}$Ge (10.37), and $^{210}$Pb (46.5), which are prominent in the open data.
  We observe a 4.3$\sigma$ local significance at 1.5~keV, near the $^{68}$Ge $L$-peak energy (1.3~keV), albeit in a region with steeply falling detection efficiency (see Fig.~\ref{fig:cuteff}).
  We observe one 3.07$\sigma$ excursion at 67.1~keV, and 30 $2\sigma$ local significance excursions, consistent with Poisson-distributed fluctuations of the spectrum.
  To determine the global significance of a signal, the local p-value is weighted by a trials factor, $p_\mathrm{global} = T\ p_\mathrm{local}$.
  $T$ typically must be computed using numerical methods, but for our data, it can be approximated as $T \approx  1+ \sqrt{\pi/2}\, \mathcal{N} Z_\mathrm{fix}$~\cite{gross2010trial}.  
  Here, $Z_\mathrm{fix}$ is the ``fixed'' desired global significance (in sigma), and $\mathcal{N} \approx 200$ is the effective number of independent search windows.
  We exclude for discovery consideration any energy $E_r$ falling at one of the 10 known lines. 
  Since no fit gives a global significance exceeding 3$\sigma$, we report only upper limits on the dark matter signal rates.

\textit{Sterile neutrino transition magnetic moment.}--
  The nonzero mass of the neutrino allows the possibility of radiative decay between states~\cite{wolfenstein1982}, including transitions of heavy right-handed sterile neutrinos $\nu_s$ into active neutrinos, $\nu_a \equiv \nu_{e, \mu, \tau}$.
  Current best limits on the magnetic moment associated with this transition come from solar neutrino-electron scattering in Borexino ($\nu_\mu \rightarrow \nu_s$)~\cite{miranda2021low}.
  Sterile neutrinos have been considered as a possible DM candidate~\cite{adhikari2017dmsterile}, and flavor-dependent couplings have been proposed to avoid constraints from SN1987A and the cosmic microwave background~\cite{shoemaker2020active, brdar2021transition}.
  
  We consider an atomic ionization process by sterile neutrinos, $\nu_s + A \rightarrow \nu_a + A^+ + e^-$.
  The limit on the 4-momentum transfer $q^2\rightarrow 0$ is kinematically accessible due to the two-body atomic final state consisting of the positive ion and ionized electron.
  Near the limit, this can be viewed as a two-step process, where the virtual exchange photon emitted from the incoming $\nu_s$ then interacts coherently with a target atom, producing ionization.
  This is known as the equivalent photon approximation (EPA)~\cite{chen2016atomicionization}.
  The singularity due to the real photon pole in the interaction cross section is accessed, enhancing it by orders of magnitude at the resonant energy $E = m_s/2$, which results in a peaked signature.
  Within the interval $E = m_s/2\ \pm\ |\vec{k}_s| / 2$, where $\vec{k}_s$ is the sterile neutrino momentum vector, the differential cross section has the form
  \begin{equation}
      \frac{d\sigma(m_s,v)}{dE} \approx \Big( \frac{\mu_{sa}}{2 m_e} \Big)^2 \frac{\alpha}{2n_A} \frac{m_s^2}{|\vec{k}_s|^2}~,
      \label{eq:tmm_peak_height}
  \end{equation}
  where $\alpha$ is the fine structure constant, $n_A$ is the number density of Ge atoms, and $\vec{k}_s$ is the momentum vector of the incoming sterile neutrino with magnetic moment $\mu_{sa}$.

\begin{figure}
    \centering
    \includegraphics[width=\columnwidth]{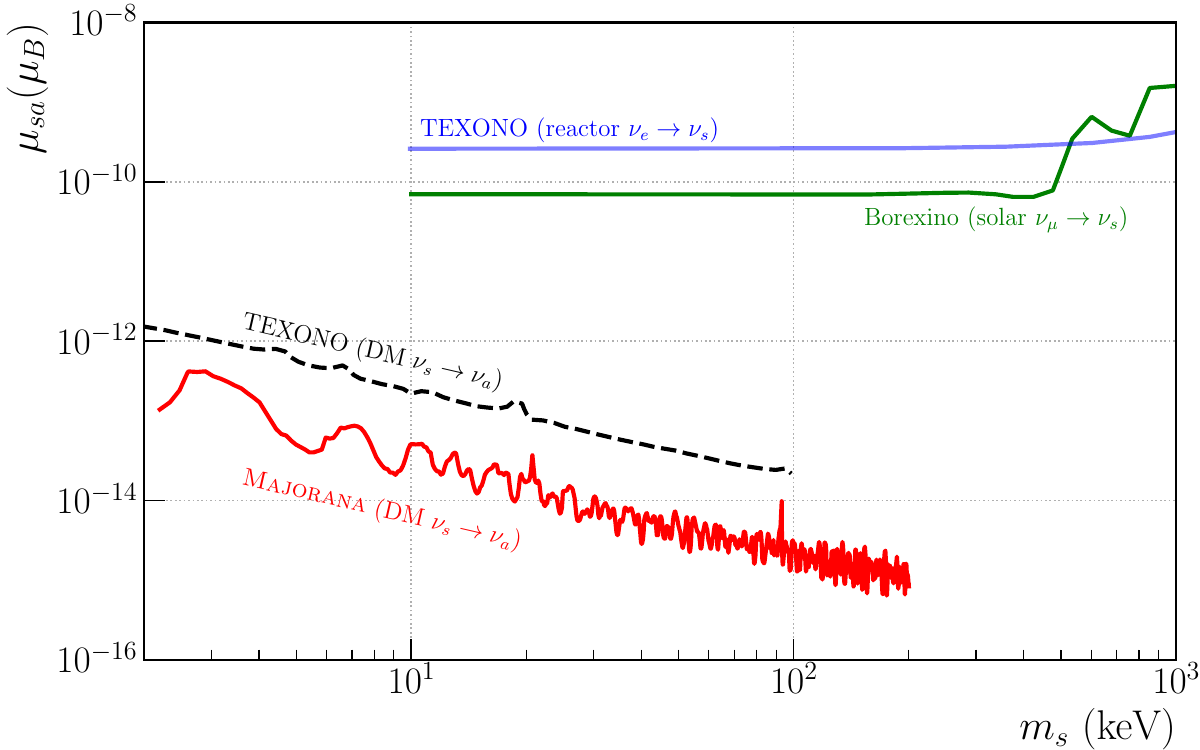}
    \caption{90\% exclusion limit on the sterile neutrino transition magnetic moment $\mu_{sa}$ (red curve, in units of the Bohr magneton $\mu_B$) via atomic ionization from \MJ\ (red), assuming sterile neutrino DM.}
    \label{fig:transition_mm_limit}
  \end{figure}

  Considering this resonance in the presence of a rich $\nu_s$ source (DM) allows more stringent (but conditional) limits to be set on $\mu_{sa}$ than previous limits from solar neutrino-electron scattering.
  Here, we assume that the local DM halo consists of $\nu_s$, for comparison with Ref.~\cite{chen2016atomicionization}
  We take the standard value for the local DM density in all models considered $\rho_{DM}$ = 0.3~GeV~cm$^{-3}$~\cite{baxter2021recommended}.
  The interaction rate in a terrestrial detector with isotopic mass $m_A$ is given by
  \begin{equation}
      \frac{dR}{dE} = \frac{\rho_\chi}{m_\chi m_A}\int^\infty_{u_{\mathrm{min}}} \Big[ \frac{d\sigma(m_s,\vec{u})}{dE}u f(\vec{u}) \Big]d^3u,
      \label{eq:DMRate}
  \end{equation}
  where $\vec{u}=\vec{v} + \vec{v_e}$ is the velocity of the dark matter (mass $m_\chi$) in the Earth's reference frame, $\vec{v}$ is the velocity of the dark matter in the galactic rest frame, $\vec{v_e}=\vec{v_e}(t)$ is the circular motion of the Earth in the galactic frame, $u_\mathrm{min}$ is the minimum speed of the DM to produce detectable recoil, and $f(\vec{u})$ is the DM velocity distribution in the Earth's reference frame.
  The EPA cross section has a resonance at $E = m_s/2$ with $u^{-2}$-dependent height and $u$-dependent width.
  All $u$-dependence in the integrand is cancelled except $f(\vec{u})$, which integrates to unity by definition.
  The expression for the event rate $R$ is thus independent of the DM velocity distribution.
  With the EPA differential cross section at the resonant energy, and setting $u_\mathrm{min} = 0$ for inelastic scattering, Eq.~\ref{eq:DMRate} can be integrated to get an expression for $R$, and set equal to $n_\mathrm{rare}/MT$, the experimental upper limit on $n_\mathrm{rare}$ at each energy,
  \begin{equation}
      \frac{n_\mathrm{rare}}{MT} = \frac{\rho_\chi\ \mu_{sa}^2\ \alpha\ m_s^2}  {m_A\ 4m_e^2\ 2n_A}.
      \label{eq:tmmrate} 
  \end{equation}
  The resulting limits on $\mu_{sa}$ are shown in Fig.~\ref{fig:transition_mm_limit}, comparing to existing $\mu_{sa}$ limits from TEXONO and Borexino~\footnote{The original $\mu_{sa}$ limit using the same postulate in Ref.~\cite{chen2016atomicionization} was found to be erroneous by a factor 14, which has been taken into account in Fig.~\ref{fig:transition_mm_limit}.
  An erratum is in preparation by the authors~\cite{chen2021private}.}.

\textit{Fermionic DM.}--
  Recent work has shown that if the DM is fermionic, its interaction with neutrinos can be modeled as Yukawa-like with a bosonic mediator~\cite{dror2020absorption}.
  In the presence of this mediator, fermionic DM can be absorbed by a nucleus, converting to neutrinos via a 2~$\rightarrow$~2 neutral-current (NC) interaction, $\chi + A \rightarrow \nu + A$~\cite{dror2020detecting, li2022fermionic}.
  For nonrelativistic DM, the signature of this inelastic scattering is a peak at the nuclear recoil energy $E_R \simeq m^2_\chi / 2M_T$, where $M_T$ is the isotopic mass of the target.
  The total absorption rate is given by
  \begin{equation}
    \frac{n_\mathrm{rare}}{MT} = \frac{\rho_\chi}{m_\chi} \sigma_\mathrm{NC} \sum_j N_{Aj} Z^2_j F_j(m_\chi)^2 \Theta(E_{R,j} - E_\mathrm{th}).
    \label{eqn:fdmabs} 
  \end{equation}
  Here, the quantity of interest is the DM-nucleon scattering cross section $\sigma_\mathrm{NC}$.

  The sum over target nuclei $j$ is performed over the five most abundant Ge isotopes in the \enrge\ detectors, weighted by the relative molecular weights of the \enrge\ detectors (75.668 $\pm$ 0.010~g/mol) to the standard weight of \natge, 72.63~g/mol.
  $N_{Aj}$ is the number of nuclei for each isotope with mass number $A_j$, and $F_j(m_\chi)$ is the normalized Helm form factor~\cite{helm1956inelastic, lewin1996review} evaluated at momentum transfer $q=m_\chi$ for non-relativistic incoming DM.
  The step function $\Theta$ is the experimental energy threshold $E_{th}$ and represents the nuclear recoil energy threshold for detectable signals.
  For the models considered in this work, we assume the cross sections are sufficiently small that the depth of the overburden does not affect the DM velocity distribution, and the probability of a multiple-scatter event is negligible. 
  
  When the dark matter signal is a nuclear recoil, the quenching factor converting from nuclear recoil energy (keVnr) to electron-equivalent (keVee or keV) energy is also considered.
  For germanium, recent work has emphasized sub-keVnr energies~\cite{collar2021quenching}.
  For our relatively higher (1--100)-keVee range, the Lindhard model with a floating $k$ parameter is more appropriate, $k = 0.16 \pm 0.02$~\cite{conus2022quenching, texono2016}.
  Intuitively, a monoenergetic nuclear recoil peak could be found at a range of different observed electron-equivalent energies $E_{ee}$, and the uncertainty in the conversion must be accounted for.
  When we set the limits on the nuclear recoil, we multiply the likelihood function $\mathcal{L}$ by a Gaussian constraint $\mathcal{L}_\mathrm{Q}$ to account for the quenching factor uncertainty.
  This tends to smooth the upper limit on the number of counts, and reduces the rare-event sensitivity in a region where a strong background peak is nearby
  \footnote{We note that the statistical analysis method to compute upper limits was selected before unblinding, but the methods to compute the local significance and the Ge quenching factor uncertainty were incorporated during the review of this manuscript.}.
  
  \begin{figure}
    \centering
    \includegraphics[width=\columnwidth]{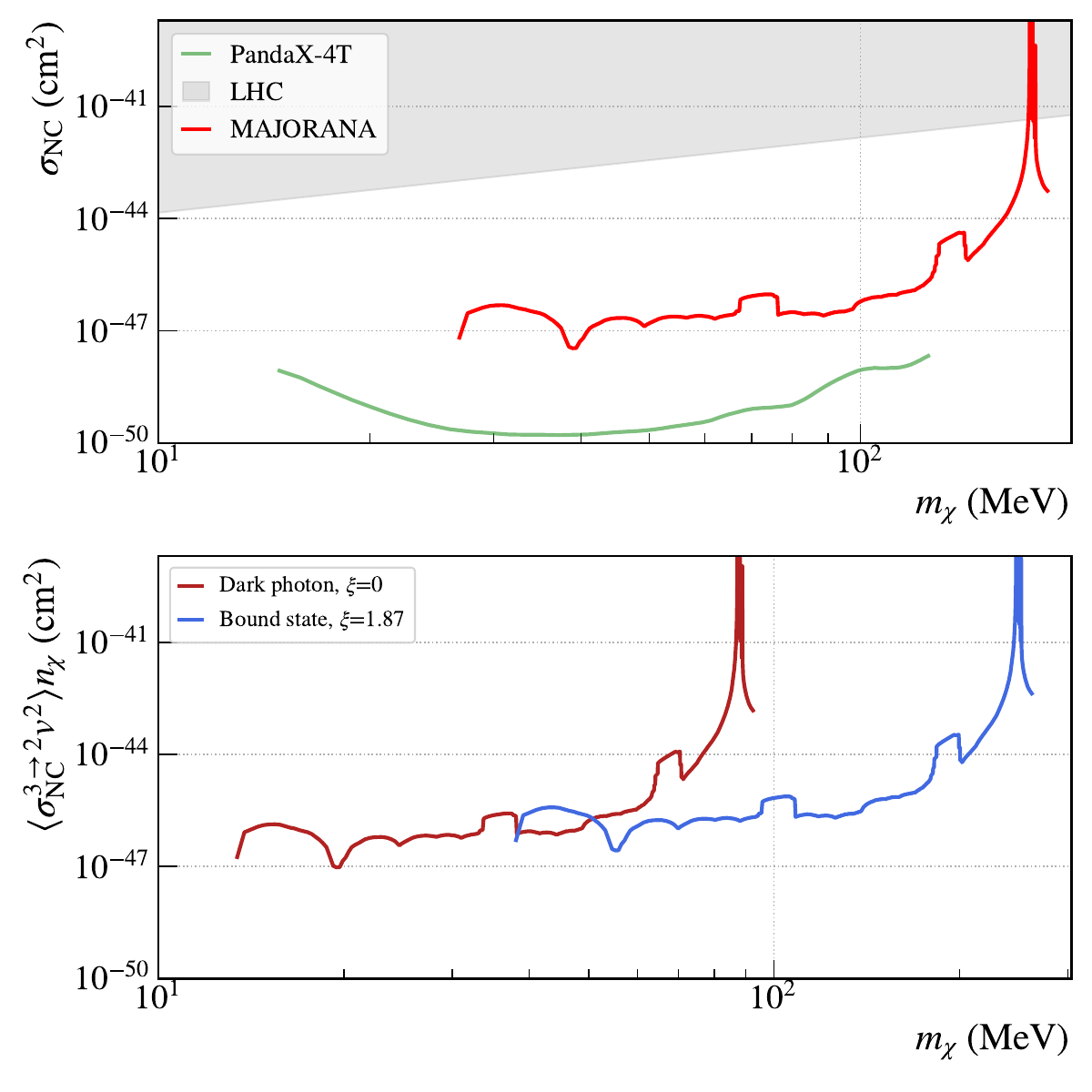}
    \caption{Top: 90\% exclusion limit for the $\chi + A \rightarrow \nu + A$ fermionic DM absorption process.
    The Helm form factor for Ge determines the overall shape of the curve, and contains a pole which produces the peak-like structure at 174~MeV~\cite{helm1956inelastic, lewin1996review}.
    Bottom: 90\% exclusion limit for DM-nucleus 3-2 inelastic scattering, $\chi + \chi + A \rightarrow \phi + A$.
    $\xi = 0$ is a massless dark photon final state and $\xi = 1.87$ is a bound final state.}
    \label{fig:fdmexc}
  \end{figure}

  Experimental search results for fermionic dark matter include $Z_0$ monojet searches at the LHC~\cite{lhc2019interplay}, and recent results from PandaX-4T with a search using 0.63 ton yr of exposure in two mass ranges~\cite{pandaX_2022, zhang2022search}.
  EXO-200 has also published a search in a lower mass range~\cite{exo200_2022}.
  Our search for fermionic dark matter, shown in Fig.~\ref{fig:fdmexc} is the first done with a Ge array surpassing the $Z_0$ monojet constraints, but is surpassed by the PandaX-4T result due to the significantly larger exposure.
  Our result is also the first to set experimental bounds for masses above 120 MeV.

\textit{Sub-GeV DM-nucleus 3$\rightarrow$2 scattering:}--
  Ref.~\cite{chao2021_dmnuc32} observed that the probability of observing DM-nucleus interactions in Ge could be significantly enhanced for sub-GeV DM if the 3$\rightarrow$2 process $\chi + \chi + A \rightarrow \phi + A$ is considered, with two DM particles in the initial state interacting coherently with the nucleus $A$.
  The signature of this process is an absorption peak at the nuclear recoil energy $E_R \simeq (4 - \xi^2) m_\chi^2 / 2 M_T$, where $\xi$ is the mass ratio of the final and initial dark matter states $\phi$ and $\chi$, and $M_T$ is the isotopic mass of the target.
  The value of $\xi$ is model-dependent, and is 0 for a massless (dark photon) final state. 
  For a bound DM final state, it is obtained by $\xi = (2m_\chi + \epsilon_1) / m_\chi$, where the binding energy is $\epsilon_1 = - (g_D^4m_\chi) / (64\pi^2)$ and $g_D$ is the new gauge coupling~\cite{petraki2015dark}.
  Setting the gauge coupling $|g_D|=3$ for the bound state DM as in Ref.~\cite{chao2021_dmnuc32}, we obtain  $\xi = 1.87$ for the bound final state.

  The total rate of nuclear recoil events has a similar form to the fermionic DM absorption [Eq.~\ref{eqn:fdmabs}]:
  \begin{align}
  \begin{split}
    \frac{n_\mathrm{rare}}{MT} &= \Big(\frac{\rho_\chi}{m_\chi}\Big)^2 \langle \sigma_\mathrm{NC}^{3\rightarrow2} v^2 \rangle \\
      &\times \sum_j N_{Aj}\ A^2_j\ F_j(q)^2\ \Theta(E_{R,j} - E_\mathrm{th}),
  \end{split}
  \end{align}
  \noindent
  where $\langle \sigma_\mathrm{NC}^{3\rightarrow2} v \rangle$ is the average three-body inelastic cross section per nucleon with the initial DM velocity $v$.

  Our search is the first to set an experimental limit for this 3 $\rightarrow$ 2 scattering process.
  We place our limit on the $(m_\chi,\sigma_\mathrm{NC}^{3\rightarrow2} v^2 n_\chi)$ parameter as suggested in Ref.~\cite{chao2021_dmnuc32}, where $n_\chi = \rho_\chi/m_\chi$ is the DM number density.
  The 90\% exclusion limits are shown in Fig.~\ref{fig:fdmexc}.

\begin{figure}
    \centering
    \includegraphics[width=\columnwidth]{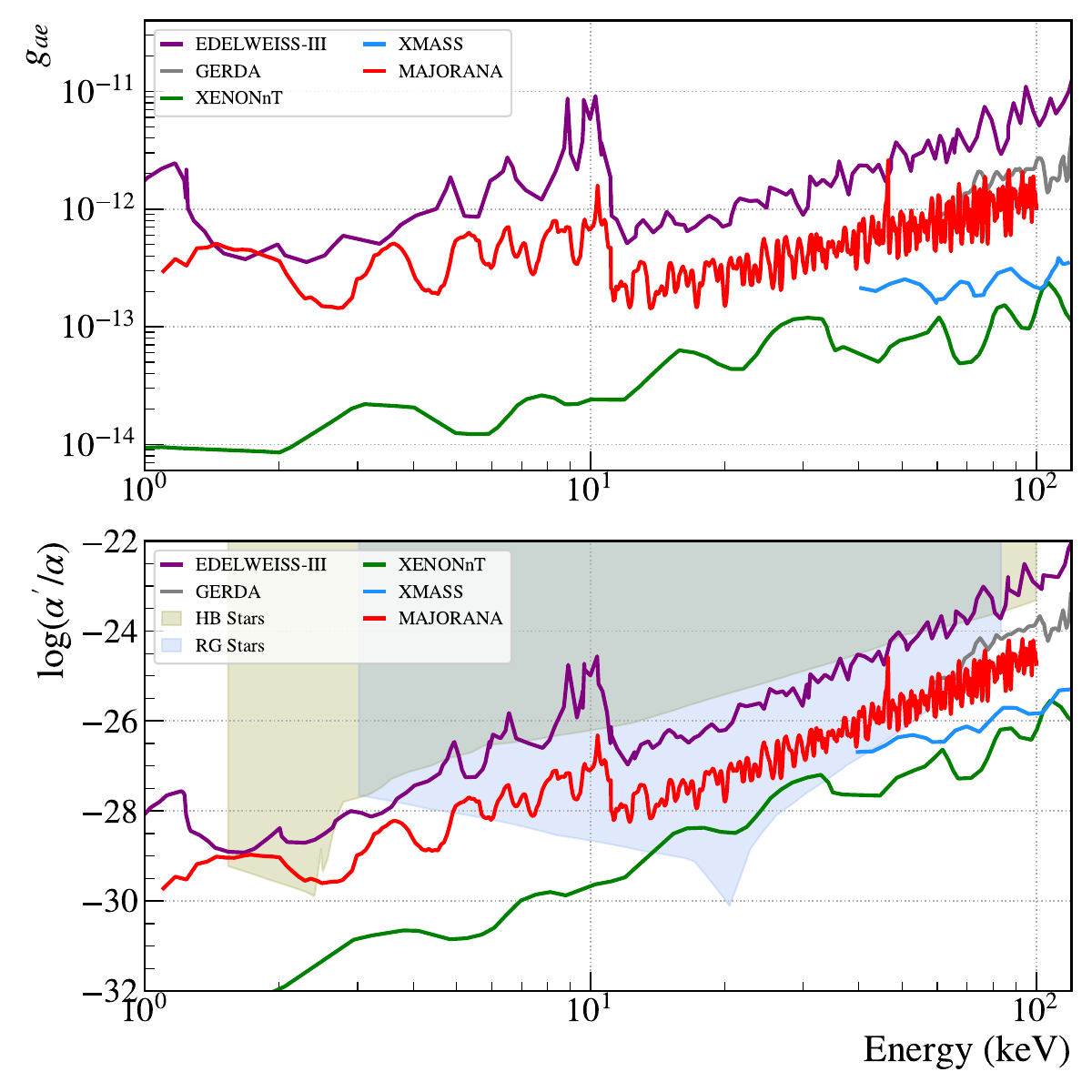}
    \caption{Top: 90\% C.L.~exclusion limits for pseudoscalar (axionlike) bosonic dark matter.  The \MJ\ result (red) is the best limit for any Ge experiment, using increased exposure and a factor 5 reduction in energy threshold from the previous analysis~\cite{vorren2017}.
    Bottom: 90\% C.L.~exclusion limits for vector bosonic (dark photon) dark matter, surpassing astrophysical limits from RG stars at the upper and lower mass range.}
    \label{fig:bdmexc}
  \end{figure}

\textit{Bosonic DM.}--
  Several experiments have searched for both pseudoscalar (axionlike) and vector (dark photon) bosonic dark matter~\cite{pospelov2008}.
  These are nonrelativistic DM candidates whose mass energy is absorbed by a target atom through a variation of the photoelectric effect~\cite{dimopoulos1985, avignone1987}, producing a peak at the rest mass energy.
  For axionlike particles, we assume the standard DM density, and a DM velocity such that the energy is approximately equal to its mass, $\beta = v_\chi / c = 0.001$, defining $m_\chi$ as the rest mass in keV.
  The DM flux (cm$^{-2}$ d$^{-1}$) becomes
  \begin{equation}
    \Phi_{\mathrm{DM}} = \frac{ \rho_{\chi}\ v_\chi}{m_\chi} = \frac{7.8 \times 10^{-17}} {m_\chi}.
  \end{equation}
  The interaction has a cross section given by~\cite{dimopoulos1985, avignone1987}:
  \begin{equation}
    \sigma_{ae}(E) = g_{ae}^2  \frac{E^2 \sigma_{pe}(E)}{\beta} \bigg(\frac{3}{16\pi\alpha m_e^2}\bigg).
  \end{equation}
  Here, $m_e$ is the electron mass in keV, $\sigma_{pe}$ is the photoelectric cross section for Ge at $E_r$~\cite{xcom2010}.
  The upper limit on the pseudoscalar coupling \gae\ can be expressed as follows, factoring it out of $\sigma_{ae}$ such that $\sigma_{ae} \equiv g_{ae}^2 \sigma_{ae}'$:
  \begin{equation}
    |g_{ae}| \leq \bigg(\frac{n_\mathrm{rare}\ m_\chi}{MT\ (7.8 \times 10^{17})\ \sigma_{ae}'(m_\chi)}\bigg)^{1/2}.
  \end{equation}
  The resulting exclusion limits are given in Fig.~\ref{fig:bdmexc}.

  Ref.~\cite{ferreira2022direct} has argued that the parameter space searched by \MJ\ and other experiments is already constrained by limits on the axionlike particle lifetime, and exclusion limits from $\gamma$- and x-ray astronomy in the range 6 keV to 1 MeV.
  In this search, we are able to exclude an additional portion of the parameter space to 1~keV, though it is also constrained by Xe experiments with larger exposures.

  For vector bosonic dark matter (dark photons), the coupling constant to electrons $\alpha'$ is related to the electromagnetic fine structure constant $\alpha$, with experimental limits set on the kinetic mixing $\kappa^2 = \alpha' / \alpha$ or its logarithm~\cite{xmass2018, pandaX_2017, armengaud2018, cdms2020, gerda2020}.
  To compute the expected counts, the product of the DM flux and interaction cross section (kg$^{-1}$ d$^{-1}$) is replaced by~\cite{pospelov2008}:
  \begin{equation}
    \Phi_{DM}\ \sigma_{ve} = \frac{4 \times 10^{23}}{m_\chi} \bigg(\frac{\alpha'}{\alpha}\bigg) \frac{\sigma_{pe}(m_\chi)}{A}.
    \label{eqn:vecdm}
  \end{equation}
  Searching for a rare peak at each energy $E_r$ as before, we obtain a limit on the coupling as a function of the mass $m_\chi$, given in Fig.~\ref{fig:bdmexc}:
  \begin{equation}
    \frac{\alpha'}{\alpha} \leq  \bigg(\frac{n_\mathrm{rare}\ A\ m_\chi}{MT\ (4 \times 10^{23})\ \sigma_{pe}(m_v)}\bigg).
  \end{equation}

\textit{Conclusions and outlook.}--
  \textsc{Majorana} has achieved the lowest background in the (1--100)-keV region of any large-scale Ge experimental search to date.
  Leveraging this low background, we have set the most stringent limits on these exotic DM models in Ge.
  While dark matter models typically do not predict a dependence on isotope, cross-checks with different isotopes would provide an important constraint on systematic errors if an experiment were to claim discovery.
  LEGEND-200~\cite{legend1000pcdr} and CDEX-300$\nu$~\cite{cdex300nu2022} face significant backgrounds in this region from $^{39}$Ar, which limits sensitivity to these signatures.
  SuperCDMS faces related challenges at lower energies~\cite{cdmsSnowmass2023}.
  Next-generation Ge arrays such as LEGEND-1000 and CDEX-1T~\cite{cdex2023} could probe new regions of parameter space if similiar background levels can be achieved.
  
\begin{acknowledgments}

  This material is based upon work supported by the U.S.~Department of Energy, Office of Science, Office of Nuclear Physics under Contracts and Awards No.~DE-AC02-05CH11231, No.~DE-AC05-00OR22725, No.~DE-AC05-76RL0130, No.~DE-FG02-97ER41020, No.~DE-FG02-97ER41033, No.~DE-FG02-97ER41041, No.~DE-SC0012612, No.~DE-SC0014445, No.~DE-SC0018060, and No.~LANLEM77, and No.~LANLEM78. 
  We acknowledge support from the Particle Astrophysics Program and Nuclear Physics Program of the National Science Foundation through Grants No.~MRI-0923142, No.~PHY-1003399, No.~PHY-1102292, No.~PHY-1206314, No.~PHY-1614611, No.~PHY-1812409, No.~PHY-1812356, and No.~PHY-2111140. 
  For this work, we gratefully acknowledge the support of the Laboratory Directed Research \& Development (LDRD) program at Lawrence Berkeley National Laboratory for this work.
  We gratefully acknowledge the support of the U.S.~Department of Energy through the Los Alamos National Laboratory LDRD Program, the Oak Ridge National Laboratory LDRD Program, and the Pacific Northwest National Laboratory LDRD program for this work.
  We gratefully acknowledge the support of the South Dakota Board of Regents Competitive Research Grant,
  We acknowledge the support of the Natural Sciences and Engineering Research Council of Canada, funding reference number SAPIN-2017-00023, and from the Canada Foundation for Innovation John R.~Evans Leaders Fund.  
  We acknowledge support from the 2020/2021 L'Or\'eal-UNESCO for Women in Science Programme.
  This research used resources provided by the Oak Ridge Leadership Computing Facility at Oak Ridge National Laboratory and by the National Energy Research Scientific Computing Center, a U.S.~Department of Energy Office of Science User Facility. 
  We thank our hosts and colleagues at the Sanford Underground Research Facility for their support.

\end{acknowledgments}

\textit{Appendix: Spectral analysis.}--
  The surface exposure of the \enrge\ detectors was carefully limited during fabrication and storage, resulting in the lowest cosmogenic activation of any Ge experiment to date and increased sensitivity to low-statistics rare events~\cite{mjdProcessing2018}.
  Since the surface exposure time of the \natge\ detectors was not limited, they show a strong tritium feature and associated cosmogenic lines, while the \enrge\ detectors show significantly reduced $^{68}$Ge, $^{65}$Zn, $^{55}$Fe, and tritium.

  The combined efficiency for \enrge\ detectors is given in Fig.~\ref{fig:cuteff}, with the centroid fit to a Weibull function.
  We show exposure weighted contributions from the time-dependent detector energy thresholds, and a flat 95\% high-frequency noise rejection efficiency.
  The fast event acceptance efficiency (or slow pulse cut) as a function of the energy is computed for each detector, rising to 95\% at 20 keV.
  It is affected by its relative position to the calibration track and available amount of small-angle Compton scatter calibration events~\cite{wiseman2019}.
  The \enrge\ spectrum shows a rising spectral shape below 10~keV, which persists after aggressive slow pulse cuts, indicating the excess signal is dominated by fast events.
  These may originate from ionization in the main fiducial (bulk) volume, events near the $p^+$ contact, or from the micrometer-thick amorphous Ge passivation layer for the PPC detector geometry, which has been observed by CDEX~\cite{cdex2022surfaceevt}.
  Despite detector storage in nitrogen environments, residual contamination of the detector component surfaces by long-lived Rn progeny including $^{210}$Pb, $^{210}$Bi, and $^{210}$Po plausibly explain the signal.
  We observe the 46.5~keV peak from $^{210}$Pb at the same intensity in both sets of detectors, but notably do not observe the associated 10.8~keV line, which may not penetrate the passivated surface region.
  Low-energy $\beta$ emission from this decay chain is also expected, which may penetrate the thin passivation layer.

\bibliographystyle{apsrev4-1}
\bibliography{refs}

\end{document}